\documentclass[twocolumn,english,superscriptaddress,citeautoscript,preprintnumbers,amsmath,amssymb,prb,floatfix,footinbib]{revtex4}
\usepackage[scaled=2]{helvet}
\usepackage[T1]{fontenc}
\usepackage[latin9]{inputenc}
\usepackage{array}
\usepackage{textcomp}
\usepackage{multirow}
\usepackage{amsmath}
\usepackage{amssymb}
\usepackage{graphicx}

\makeatletter

\providecommand{\tabularnewline}{\\}

\@ifundefined{textcolor}{}
{%
 \definecolor{BLACK}{gray}{0}
 \definecolor{WHITE}{gray}{1}
 \definecolor{RED}{rgb}{1,0,0}
 \definecolor{GREEN}{rgb}{0,1,0}
 \definecolor{BLUE}{rgb}{0,0,1}
 \definecolor{CYAN}{cmyk}{1,0,0,0}
 \definecolor{MAGENTA}{cmyk}{0,1,0,0}
 \definecolor{YELLOW}{cmyk}{0,0,1,0}
}

\usepackage[scaled=2]{helvet}\usepackage{amstext}

\makeatletter
\@ifundefined{textcolor}{}{\definecolor{BLACK}{gray}{0}\definecolor{WHITE}{gray}{1}\definecolor{RED}{rgb}{1,0,0}\definecolor{GREEN}{rgb}{0,1,0}\definecolor{BLUE}{rgb}{0,0,1}\definecolor{CYAN}{cmyk}{1,0,0,0}\definecolor{MAGENTA}{cmyk}{0,1,0,0}\definecolor{YELLOW}{cmyk}{0,0,1,0}}

\usepackage{dcolumn}
\usepackage{bm}
\usepackage{times}
\usepackage{hyperref}
\hypersetup{colorlinks=true,linkcolor=red,citecolor=blue}

\makeatother

\usepackage{babel}

\makeatother

\usepackage{babel}
\begin{document}
\title{Evolution from helical to collinear ferromagnetic order of the Eu$^{2+}$ spins in RbEu(Fe$_{1-x}$Ni$_{x}$)$_{4}$As$_{4}$}
\author{Qianhui Xu}
\thanks{These authors contributed equally to this work.}
\affiliation{School of Physics, Beihang University, Beijing 100191, China}
\author{Yi Liu}
\thanks{These authors contributed equally to this work.}
\affiliation{College of Science, Zhejiang University of Technology, Hangzhou 310023, China}
\affiliation{Department of Physics, Zhejiang University, Hangzhou 310027, China}
\author{Sijie Hao}
\thanks{These authors contributed equally to this work.}
\affiliation{Department of Physics, Beijing Normal University, Beijing 100875, China}
\author{Jiahui Qian}
\affiliation{State Key Laboratory of Surface Physics and Department of Physics, Fudan University, Shanghai 200433, China}
\author{Cheng Su}
\affiliation{School of Physics, Beihang University, Beijing 100191, China}
\author{Chin-Wei Wang}
\affiliation{National Synchrotron Radiation Research Center, Hsinchu, 30077, Taiwan}
\author{Thomas Hansen}
\affiliation{Institut Laue-Langevin, BP 156, 38042 Grenoble Cedex 9, France}
\author{Zhendong Fu}
\affiliation{Neutron Platform, Songshan Lake Materials Laboratory, Dongguan 523808, China}
\author{Yixi Su}
\affiliation{J\"{u}lich Centre for Neutron Science JCNS at Heinz Maier-Leibnitz Zentrum (MLZ), Forschungszentrum J\"{u}lich GmbH, Lichtenbergstra{\ss}e 1, D-85747 Garching, Germany}
\author{Wei Li}
\affiliation{State Key Laboratory of Surface Physics and Department of Physics, Fudan University, Shanghai 200433, China}
\author{Guang-Han Cao}
\email{ghcao@zju.edu.cn}
\affiliation{Department of Physics, Zhejiang University, Hangzhou 310027, China}
\author{Yinguo Xiao}
\email{y.xiao@pku.edu.cn}
\affiliation{School of Advanced Materials, Peking University Shenzhen Graduate School, Shenzhen 518055, China}
\author{Wentao Jin}
\email{wtjin@buaa.edu.cn}
\affiliation{School of Physics, Beihang University, Beijing 100191, China}

\begin{abstract}
The ground-state magnetic structures of the Eu$^{2+}$ spins in recently discovered RbEu(Fe$_{1-x}$Ni$_{x}$)$_{4}$As$_{4}$ superconductors have been investigated by neutron powder diffraction measurements. It is found that as the superconductivity gets suppressed with the increase of Ni doping, the magnetic propagation vector of the Eu sublattice diminishes, corresponding to the decrease of the rotation angle between the moments in neighboring Eu layers. The ferromagnetic Eu layers are helically modulated along the $\mathit{c}$ axis with an incommensurate magnetic propagation vector in both the ferromagnetic superconductor RbEu(Fe$_{0.95}$Ni$_{0.05}$)$_{4}$As$_{4}$ and the superconducting ferromagnet RbEu(Fe$_{0.93}$Ni$_{0.07}$)$_{4}$As$_{4}$. Such a helical structure transforms into a purely collinear ferromagnetic structure for non-superconducting RbEu(Fe$_{0.91}$Ni$_{0.09}$)$_{4}$As$_{4}$, with all the Eu$^{2+}$ spins lying along the tetragonal (1 1 0) direction. The evolution from helical to collinear ferromagnetic order of the Eu$^{2+}$ spins with increasing Ni doping is supported by first-principles calculations. The variation of the rotation angle between adjacent Eu$^{2+}$ layers can be well explained by considering the change of magnetic exchange couplings mediated by the indirect Ruderman-Kittel-Kasuya-Yosida (RKKY) interaction.
\end{abstract}
\maketitle

\section{Introduction}

The discovery of iron-based superconductors in 2008 has stimulated
worldwide research interests in the investigations of the interplay
between magnetism and unconventional superconductivity in these novel
materials.\cite{Kamihara_08,Dai_15} Among various members of the
iron-based superconductors, the ternary \textquotedblleft EuFe$_{2}$As$_{2}$\textquotedblright{}
(Eu122) system is a unique representative and has attracted much attention,
due to the existence of two magnetic sublattices in the unit cell
and the strong coupling between spin-, lattice- and charge degrees
of freedom.\cite{Zapf_17,Xiao_10,Xiao_12} The undoped parent compound
EuFe$_{2}$As$_{2}$ shows an $\mathit{A}$-type antiferromagnetic
(AFM) order of the localized Eu$^{2+}$ spins below 19 K, in addition
to the spin-density-wave (SDW) order of the itinerant Fe moments below
190 K.\cite{Ren_08,Xiao_09} By suppressing the SDW order in the Fe
sublattice, superconductivity can be achieved by means of chemical
substitutions or applying external pressure.\cite{Zapf_17,Ren_09,Miclea_09}
In the superconducting ground state, single-crystal neutron diffraction
or x-ray resonant magnetic scattering experiments have confirmed that
strong ferromagnetism from Eu 4$\mathit{f}$ orbitals with an ordered
moment of $\sim$ 7 $\mathit{\mu_{B}}$ per Eu atom can coexist microscopically
with bulk superconductivity and reach a compromise.\cite{Jin_13,Nandi_14,Jin_15,Jin_Ru,Jin_Pressure}
The intriguing coexistence of ferromagnetism and superconductivity
revealed in the Eu122 system drives the experimental efforts to further
explore other novel Eu-containing iron-based superconductors.

In 2016, superconductivity with the transition temperature ($\mathit{T_{SC}}$)
of approximately 31-36 K was discovered in a new family of iron pnictides
Ca$\mathit{A}$Fe$_{4}$As$_{4}$ and Sr$\mathit{A}$Fe$_{4}$As$_{4}$
($\mathit{A}$ = K, Rb, Cs) possessing the \textquotedblleft 1144\textquotedblright -type
structure.\cite{Iyo_16} Later on, RbEuFe$_{4}$As$_{4}$ (denoted
as Eu1144 below), crystallizing as an intergrowth structure of heavily
hole-doped superconducting RbFe$_{2}$As$_{2}$ ($\mathit{T_{SC}}$
= 2.6 K)\cite{Bukowski_10} and non-superconducting EuFe$_{2}$As$_{2}$,
was reported to be a superconductor as well with $\mathit{T_{SC}}$
= 36 K.\cite{Liu_16,Smylie_18} The FeAs layers in Eu1144 are intrinsically
hole doped due to the charge homogenization associated with the structural
hybridization, which is responsible for the absence of Fe-SDW order
and the occurrence of superconductivity. Ascribing to the longer interlayer
distance between the Eu layers in Eu1144 compared to Eu122, the Eu$^{2+}$
spins order magnetically at a lower temperature of $\mathit{T_{m}}$
= 15 K. 

Based on the magnetization and specific heat data obtained from high-quality
powder samples, RbEuFe$_{4}$As$_{4}$ was speculated to be a ferromagnetic
superconductor with a robust coexistence of superconductivity and
ferromagnetism.\cite{Liu_16} Neutron diffraction measurements on
a Eu1144 single crystal have been performed to clarify how the two-dimensional
in-plane ferromagnetic Eu layers stack along the $\mathit{c}$ axis.\cite{Iida_19}
A magnetic propagation vector of $\mathit{k}$ = (0, 0, 0.25) is revealed,
suggesting the rotation angle of 90$^{\circ}$ between the in-plane ferromagnetically
aligned Eu$^{2+}$ spins on adjacent layers. Such a helical magnetic
structure of undoped Eu1144 is in stark contrast to the collinear A-type
AFM structure of undoped EuFe$_{2}$As$_{2}$, but resembles those
of EuCo$_{2}$As$_{2}$ and EuNi$_{2}$As$_{2}$, showing an incommensurate
magnetic propagation vector of $\mathit{k}$ = (0, 0, 0.79) and $\mathit{k}$
= (0, 0, 0.92), respectively.\cite{Tan_16,Jin_EuNi2As2} 

By introducing extra itinerant electrons via the substitution of Ni$^{2+}$
(3$\mathit{d}$$^{8}$) for Fe$^{2+}$ (3$\mathit{d}$$^{6}$), the
intrinsically doped hole carriers in RbEuFe$_{4}$As$_{4}$ can be
compensated. Systematic macroscopic characterizations including resistivity,
magnetization, and specific heat measurements have been performed
on polycrystalline and single-crystal samples of RbEu(Fe$_{1-x}$Ni$_{x}$)$_{4}$As$_{4}$
to establish the superconducting and magnetic phase diagram.\cite{Liu_17,Willa_20}
It is figured out that $\mathit{T_{SC}}$ descreases rapidly with
the Ni doping, while the magnetic ordering temperature of the Eu sublattice,
$\mathit{T_{m}}$, remains essentially unchanged. Consequently, RbEu(Fe$_{1-x}$Ni$_{x}$)$_{4}$As$_{4}$
transforms from the ferromagnetic superconductor (FSC) with $\mathit{T_{SC}}$
$>$ $\mathit{T_{m}}$ for $\mathit{x}<$ 0.07, to the so-called ``superconducting
ferromagnet'' (SFM) with $\mathit{T_{m}}$ $>$ $\mathit{T_{SC}}$
for 0.07 $\leqslant x\leqslant$ 0.08, and finally to the ferromagnetic
non-superconductor for $\mathit{x}>$ 0.09. Furthermore, a recovered
Fe-AFM state is proposed for 0.04 $\leqslant x\leqslant$ 0.10 based
on the resistivity data on polycrystalline samples.\cite{Liu_17} 

As the helical magnetic order of the Eu$^{2+}$ spins with a two-dimensional
(2D) character in undoped Eu1144 is proposed to be associated with
the presence of superconductivity,\cite{Devizorova_19,Koshelev_19}
it is of great interest to clarify how the magnetic structure of RbEu(Fe$_{1-x}$Ni$_{x}$)$_{4}$As$_{4}$
develops against the weakening of the superconductivity induced by
Ni doping. Fitting to the magnetic suceptibility in the paramagnetic
state yields comparable positive values of Currie-Weiss temperature
for samples with different $\mathit{x}$,\cite{Liu_17,Willa_20} reflecting
dominant in-plane ferromagnetic interactions between the Eu$^{2+}$
moments. Detailed neutron diffraction measurements on RbEu(Fe$_{1-x}$Ni$_{x}$)$_{4}$As$_{4}$
will deliver important information regarding how the stacking pattern
of the ferromagnetic Eu layer along the $\mathit{c}$ axis changes
with $\mathit{x}$ and how it is correlated with the suppression of
superconductivity.

Here we present a systematic study of the magnetic structures of Ni-doped
Eu1144 with different doping levels as determined by neutron powder
diffraction. We find that as the superconductivity gets suppressed
gradually with the increase of Ni doping, the magnetic propagation
vector of the Eu sublattice diminishes, corresponding to the decrease
of the rotation angle between the moments in neighboring Eu layers.
No evidence of the proposed recovery of Fe-SDW order is observed within
our experimental resolution.  The variation of the rotation angle between adjacent Eu$^{2+}$ layers can be well explained by considering the change of magnetic exchange couplings mediated by the indirect Ruderman-Kittel-Kasuya-Yosida (RKKY) interaction.

\section{Experimental Details and Calculation Methods}

Polycrystalline samples of RbEu(Fe$_{1-x}$Ni$_{x}$)$_{4}$As$_{4}$
($\mathit{x}$ = 0.05, 0.07 and 0.09) of $\sim$ 4 g were synthesized
by the solid-state reaction method as described in Ref. \onlinecite{Liu_17}.
The phase purity was checked by x-ray diffraction (XRD) on a PANalytical
x-ray diffractometer with a monochromatic Cu-K$_{\alpha1}$ radiation. The doping concentration of Ni in three samples was checked by energy-dispersive x-ray spectroscopy (EDS), to be 5.6(6) \%, 7.1(8) \%, 8.9(5) \%, respectively, well consistent with the nominal values. A small amount of FeAs impurity was found to exist in the samples
with $\mathit{x}$ = 0.05 and 0.07, and small amounts of RbFe$_{2}$As$_{2}$
and EuFe$_{2}$As$_{2}$ impurities were identified in the sample
with $\mathit{x}$ = 0.09. Low-temperature neutron powder diffraction
(NPD) measurements on the samples with $\mathit{x}$ = 0.05 were performed
on the high-intensity powder diffractometer Wombat\cite{Wombat} at
the OPAL facility (Lucas Height, Australia) using incident neutrons
with the wavelength of 2.41 \AA{} and 1.54 \AA{}, while the data of the sample
with $\mathit{x}$ = 0.07 were collected using the 1.54 \AA{} wavelength
only. NPD measurements on the sample with $\mathit{x}$ = 0.09 were
performed on the high-intensity powder diffractometer D20 at Institut
Laue-Langevin (Grenoble, France) using incident neutrons with the
wavelength of 2.41 \AA{} and 1.30 \AA{}. In order to minimize the effect of
neutron absorption by the Eu atoms, we have filled the powder samples
into the double-wall hollow vanadium cylinder. Refinements of both
nuclear and magnetic structures were carried out using the FullProf
program suite.\cite{Rodriguez-Carvajal_93}

The first-principles calculations presented in this paper are performed using the projected augmented-wave method,\cite{Blochl} as implemented in the VASP code.\cite{Kresse} The exchange correlation potential is calculated using the generalized gradient approximation (GGA) as proposed by Perdew, Burke, and Ernzerhof.\cite{Perdew} We have included the strong Coulomb repulsion in the Eu-4$\mathit{f}$ orbitals on a mean-field level using the GGA+$\mathit{U_{eff}}$ approximation. Since there exist no spectroscopy data for RbEu(Fe$_{1-x}$Ni$_{x}$)$_{4}$As$_{4}$, we have used a $\mathit{U_{eff}}$ of 8 eV throughout this work, which is the standard value for an Eu$^{2+}$ ion.\cite{LiW,Jin_15,Jin_Ru} The results have been checked for consistency with varying $\mathit{U_{eff}}$ values. $\mathit{U_{eff}}$ is not applied to the itinerant Fe-3$\mathit{d}$ and Ni-3$\mathit{d}$ orbitals. Additionally, the spin-orbit coupling is included for all atoms with the second variational method in the calculations. These calculations are performed using the experimental crystal structure, as determined by the neutron diffraction measurements.

\section{Results }

\begin{figure*}
\centering{}\includegraphics[width=1\textwidth]{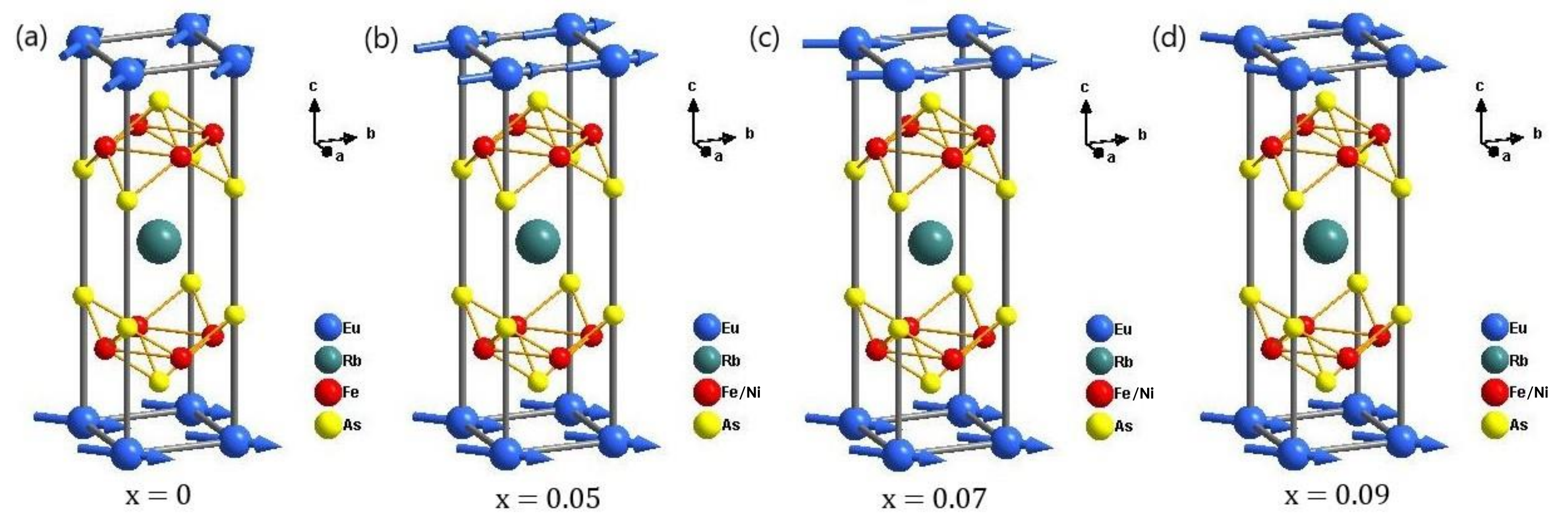}

\caption{The ground-state magnetic structure of RbEu(Fe$_{1-x}$Ni$_{x}$)$_{4}$As$_{4}$
with $\mathit{x}$ = 0 (a),\cite{Iida_19} $\mathit{x}$ = 0.05 (b),
$\mathit{x}$ = 0.07 (c), and $\mathit{x}$ = 0.09 (d), in which the
rotation angle between the in-plane ferromagnetically aligned Eu$^{2+}$
moments on adjacent layers are 90$^{\circ}$, $\sim$ 49$^{\circ}$, $\sim$ 26$^{\circ}$, and
0$^{\circ}$, respectively.}
\end{figure*}

The ground-state magnetic structures of the Eu$^{2+}$ spins in RbEu(Fe$_{1-x}$Ni$_{x}$)$_{4}$As$_{4}$
with different Ni doping levels ($\mathit{x}$ = 0.05, 0.07 and 0.09)
are determined by NPD measurements and illustrated in Figure 1(b,
c, d), together with the helical magnetic structure of undoped RbEuFe$_{4}$As$_{4}$
($\mathit{x}$ = 0) with $\mathit{k}$ = (0, 0, 0.25) (Fig. 1(a))
as determined in Ref. \onlinecite{Iida_19} , which will be discussed
in detail below.

\begin{figure*}
\centering{}\includegraphics[width=1\textwidth]{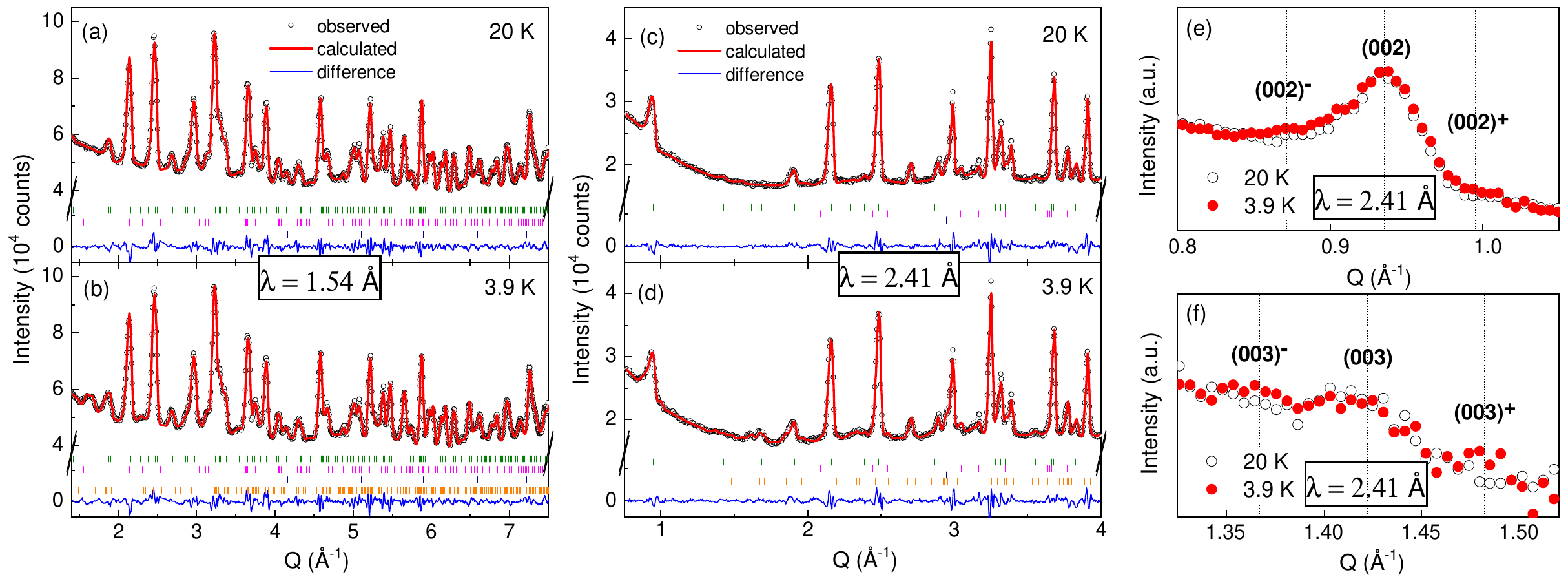}

\caption{NPD patterns of RbEu(Fe$_{0.95}$Ni$_{0.05}$)$_{4}$As$_{4}$ at
20 K (a, c) and 3.9 K (b, d) and the Rietveld refinments. The left
(a, b) and right (c-f) panels show the data collected using the incident
neutron wavelength of 1.54 \AA{} and 2.41 \AA{}, respectively. The patterns
in (b) and (d) are the refinement results obtained by adopting a magnetic
structure model with the irreducible representation $\Gamma_{5}$
as described in the text. The circles represent the observed intensities,
and the solid lines are the calculated patterns. The differences between
the observed and calculated intensities are shown at the bottom. The
vertical bars in olive, magenta, navy and orange colors indicate the
expected nuclear Bragg reflections from the RbEu(Fe$_{0.95}$Ni$_{0.05}$)$_{4}$As$_{4}$
main phase, FeAs impurity, vanadium sample container and the magnetic
Bragg reflections from RbEu(Fe$_{0.95}$Ni$_{0.05}$)$_{4}$As$_{4}$,
respectively. (e) and (f) show the  enlarged high-resolution diffraction patterns
at 3.9 and 20 K around the (0 0 2) and (0 0 3) nuclear peak positions, respectively,
visualizing the incommensurate magnetic satellite reflections appearing
at 3.9 K.}
\end{figure*}

Figure 2 shows the NPD patterns of RbEu(Fe$_{0.95}$Ni$_{0.05}$)$_{4}$As$_{4}$
at 20 K and 3.9 K. According to the superconducting and magnetic phase
diagram of RbEu(Fe$_{1-x}$Ni$_{x}$)$_{4}$As$_{4}$ deduced from
macroscopic measurements in Ref. \onlinecite{Liu_17}, for this composition,
the temperature of 20 K is above $\mathit{T_{m}}$(= 15 K) but below
$\mathit{T_{SDW}}$ (= 28.9 K), which is the SDW ordering temperature
of Fe. As shown in Fig. 2(a) and 2(c), the diffraction patterns at
20 K can be well fitted with the crystal structure reported in Ref. \onlinecite{Liu_17}
(space group $\mathit{P}$4$\mathit{/m}$$\mathit{m}$$\mathit{m}$)
with a small amount of FeAs impurity (7\% wt). Within our experimental
uncertainty, no magnetic reflections at (0.5, 0.5, 3) ($\mathit{Q}$
= 1.84 \AA{}$^{-1}$) associated with possible Fe-AFM order can be identified,
assuming that the Fe$^{2+}$ moments order in the hedgehog spin-vortex
crystal (SVC) motif in each Fe plane and are antiferromagnetically
stacked along the $\mathit{c}$ direction, similar to that observed
in isostructural CaK(Fe$_{1-x}$Ni$_{x}$)$_{4}$As$_{4}$.\cite{Meier_18,Kreyssig_18} 

Upon cooling down to 3.9 K, which is well below $\mathit{T_{m}}$,
the magnetic reflections due to the magnetic ordering of Eu appear
as satellite peaks close to the nuclear reflections. As shown in Fig.
2(d), the incident neutron wavelength of $\lambda$ = 2.41 \AA{} provides
a better resolution at low-$\mathit{Q}$ region, confirming the incommensurate
nature of the magnetic peaks. The magnetic reflections at (0 0 2)$^{-}$/(0
0 2)$^{+}$ and (0 0 3)$^{-}$/(0 0 3)$^{+}$ emerge in both sides
of the (0 0 2) and (0 0 3) peaks, as illustrated in Fig. 1(e) and 1(f). Using the k\_search program integrated in the FullProf
suite, the magnetic propagation vector of $\mathit{k}$ = (0, 0, 0.136(4))
is figured out for RbEu(Fe$_{0.95}$Ni$_{0.05}$)$_{4}$As$_{4}$. 

\begin{figure*}
\centering{}\includegraphics[width=1\textwidth]{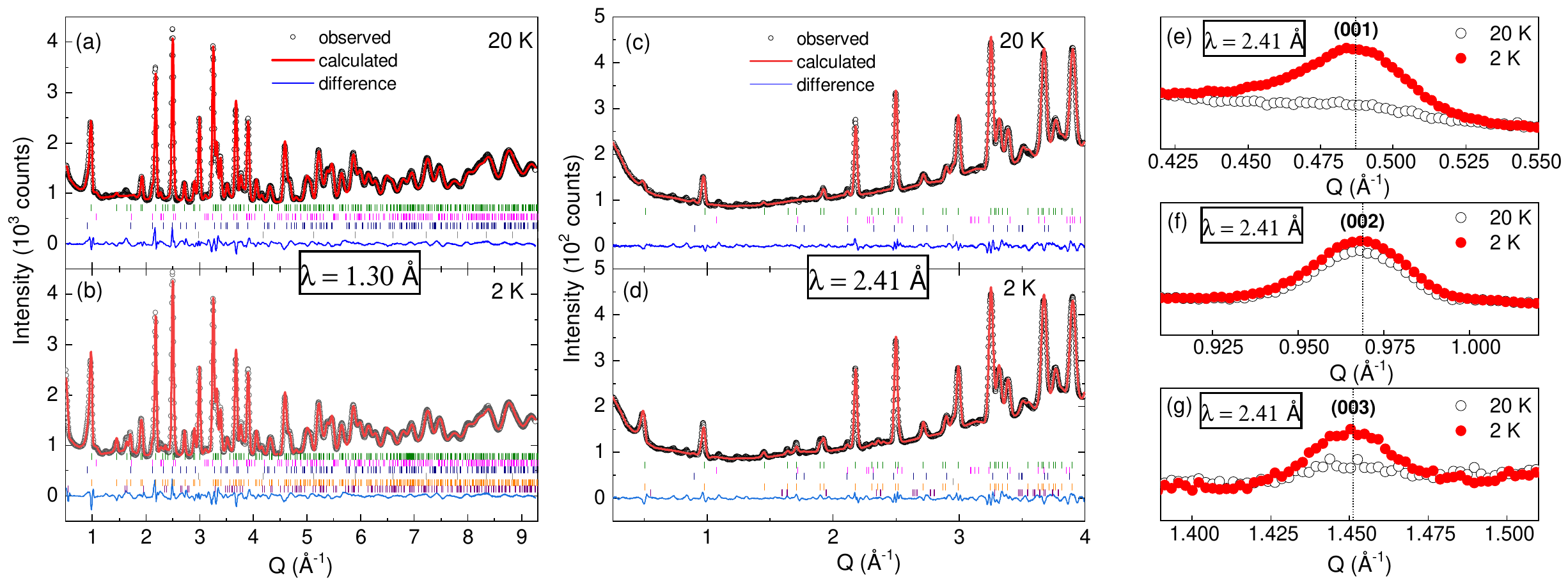}

\caption{NPD patterns of RbEu(Fe$_{0.91}$Ni$_{0.09}$)$_{4}$As$_{4}$ at
20 K (a, c) and 2 K (b, d) and the Rietveld refinments. The left (a,
b) and right (c-g) panels show the data collected using the incident
neutron wavelength of 1.30 \AA{} and 2.41 \AA{}, respectively. The patterns
in (b) and (d) are the refinement results obtained by adopting a magnetic
structure model with the irreducible representation $\Gamma_{9}$
as described in the text. The circles represent the observed intensities,
and the solid lines are the calculated patterns. The differences between
the observed and calculated intensities are shown at the bottom. The
vertical bars in olive, magenta, navy, gray, orange and purple colors indicate
the expected nuclear Bragg reflections from the RbEu(Fe$_{0.91}$Ni$_{0.09}$)$_{4}$As$_{4}$
main phase, EuFe$_{2}$As$_{2}$ impurity, RbFe$_{2}$As$_{2}$ impurity,
vanadium sample container, as well as the magnetic Bragg reflections from the 
RbEu(Fe$_{0.91}$Ni$_{0.09}$)$_{4}$As$_{4}$ main phase and the EuFe$_{2}$As$_{2}$ impurity, respectively. (e), (f) and (g) show the enlarged high-resolution diffraction patterns at 2 and 20
K around the (0 0 1), (0 0 2) and (0 0 3) nuclear peak positions, respectively, 
illustrating the commensurate magnetic contributions with $\mathit{k}$ = 0 at
2 K.}
\end{figure*}

According to the representation analysis performed using the BasIreps
program also integrated in the FullProf suite (see the supplemental materials for details), for the space group
of $\mathit{P}$4$\mathit{/m}$$\mathit{m}$$\mathit{m}$, only two
magnetic representations are possible for the Eu (1$\mathit{a}$)
site with the propagation vector of $\mathit{k}$ = (0, 0, 0.136(4)),
which we label as $\Gamma_{1}$ and $\Gamma_{5}$, respectively. $\Gamma_{1}$
allows the $\mathit{c}$-axis aligned ferromagnetic Eu layers stacking
with modulated moment size values at different layers, which is not
consistent with the easy-plane magnetization as revealed from the single-crystal sample with a similar Ni doping level.\cite{Willa_20} On the other
hand, $\Gamma_{5}$ allows the in-plane aligned ferromagnetic Eu layers
to stack helically along the $\mathit{c}$ axis, with a constant moment
size value at different layers. This model fits pretty well to the
diffraction patterns at 3.9 K, as shown by the solid curves in Fig.
2(b) and 2(d). As a comparison between the fitting using $\Gamma_{5}$ and $\Gamma_{1}$, Fig. S2 in the supplemental materials shows a better agreement of $\Gamma_{5}$ with the observed intensities in the very low-Q region, where the magnetic form factor dominates. The nuclear structure parameters and the scale factor
derived from the refinement of 20 K data was fixed in the refinement
of 3.9 K data to derive the moment size of Eu to be 6.3(2) $\mathit{\mu_{B}}$,
as listed in Table 1. As illustrated in Fig. 1(b), the Eu$^{2+}$
moments form an incommensurate helical structure, with the moment
direction lying in the $\mathit{ab}$ plane but rotating by $\sim$
49$^{\circ}$ around the $\mathit{c}$ axis with respect to adjacent Eu layers. Using the Bilbao Crystallographic Server,\cite{Bilbao} the magnetic space group of this helical structure is determined to be $\mathit{Pm'm'm}$ (No. 47.252).

\begin{table*}
\caption{Refined results for the nuclear and magnetic structure parameters
of RbEu(Fe$_{1-x}$Ni$_{x}$)$_{4}$As$_{4}$ with $\mathit{x}$ =
0.05, 0.07 and 0.09. The atomic positions are as follows: Eu, $1a$
(0, 0, 0); Rb, $1d$ (0.5, 0.5, 0.5); Fe/Ni, $4i$ (0, 0.5, $\mathit{z_{\mathrm{Fe}}}$);
As1, $2g$ (0, 0, $\mathit{z_{\mathrm{As1}}}$); As2, $2h$ (0.5,
0.5, $\mathit{z_{\mathrm{As2}}}$). The occupancies of Fe and Ni were
fixed according to the nominal compositions, respectively. The nuclear
structure parameters and the scale factor derived from the refinement
of 20 K data was fixed in the magnetic-structure refinements (Space
group: $\mathit{P}$4$\mathit{/m}$$\mathit{m}$$\mathit{m}$)}

\begin{ruledtabular} %
\begin{tabular}{cc|cc|cc|cc}
\multicolumn{2}{c|}{Composition} & \multicolumn{2}{c|}{RbEu(Fe$_{0.95}$Ni$_{0.05}$)$_{4}$As$_{4}$} & \multicolumn{2}{c|}{RbEu(Fe$_{0.93}$Ni$_{0.07}$)$_{4}$As$_{4}$} & \multicolumn{2}{c}{RbEu(Fe$_{0.91}$Ni$_{0.09}$)$_{4}$As$_{4}$}\tabularnewline
\hline 
\multicolumn{2}{c|}{Temperature} & 20 K & 3.9 K & 20 K & 3.3 K & 20 K & 2 K\tabularnewline
\hline 
\multirow{2}{*}{Eu} & $B_{iso}$ (\AA{}$^{2}$) & 1.3(1) & - & 1.2(1) & - & 0.22(5) & -\tabularnewline
 & $M$ ($\mu_{B}$) & - & 6.3(2) & - & 6.3(2) & - & 6.5(1)\tabularnewline
\hline 
Rb & $B_{iso}$ (\AA{}$^{2}$) & 1.4(1) & - & 1.3(1) & - & 1.1(1) & -\tabularnewline
\hline 
\multirow{2}{*}{Fe/Ni} & $\mathit{z_{\mathrm{Fe}}}$ & 0.2309(2) & - & 0.2310(2) & - & 0.2315(1) & -\tabularnewline
 & $B_{iso}$ (\AA{}$^{2}$) & 1.0(1) & - & 0.8(1) & - & 0.26(1) & -\tabularnewline
\hline 
\multirow{2}{*}{As1} & $\mathit{z_{\mathrm{As1}}}$ & 0.3344(4) & - & 0.3339(4) & - & 0.3339(2) & -\tabularnewline
 & $B_{iso}$ (\AA{}$^{2}$) & 1.1(1) & - & 0.8(1) & - & 0.24(3) & -\tabularnewline
\hline 
\multirow{2}{*}{As2} & $\mathit{z_{\mathrm{As2}}}$ & 0.1263(4) & - & 0.1263(4) & - & 0.1277(2) & -\tabularnewline
 & $B_{iso}$ (\AA{}$^{2}$) & 1.3(1) & - & 0.8(1) & - & 0.24(3) & -\tabularnewline
\hline 
\multicolumn{2}{c|}{$\mathit{a}$ (\AA{})} & 3.8652(4) & 3.8651(2) & 3.8649(5) & 3.8646(2) & 3.8921(3) & 3.8920(2)\tabularnewline
\multicolumn{2}{c|}{$\mathit{c}$ (\AA{})} & 13.117(2) & 13.117(1) & 13.109(2) & 13.108(1) & 13.218(1) & 13.216(1)\tabularnewline
\hline 
\multicolumn{2}{c|}{$R_{F^{2}}$} & 1.29 & 1.28 & 1.31 & 1.34 & 1.86 & 2.01\tabularnewline
\multicolumn{2}{c|}{$R_{wF^{2}}$} & 1.73 & 1.72 & 1.83 & 1.86 & 2.40 & 2.63\tabularnewline
\multicolumn{2}{c|}{$R_{F}$} & 0.43 & 0.44 & 0.42 & 0.43 & 0.36 & 0.36\tabularnewline
\end{tabular}\end{ruledtabular}
\end{table*}

Figure 3 shows the NPD patterns of RbEu(Fe$_{0.91}$Ni$_{0.09}$)$_{4}$As$_{4}$
at 20 K and 2 K. This sample is non-superconducting as evidenced from
previous macroscopic characterizations.\cite{Liu_17} It undergoes
the magnetic ordering of Eu sublattice at $\mathit{T_{m}}$ (= 14.7
K) and a possible recovered Fe-SDW ordering at $\mathit{T_{SDW}}$
(= 31.3 K). Similar to the case of $\mathit{x}$ = 0.5 presented above,
no visible change of intensities at (0.5, 0.5, 1) ($\mathit{Q}$ =
1.25 \AA{}$^{-1}$) and (0.5, 0.5, 3) ($\mathit{Q}$ = 1.84 \AA{}$^{-1}$)
associated with the Fe-AFM order can be resolved at 20 K compared
with 40 K (data of which is not shown). The diffraction patterns at
20 K can be well fitted using the nuclear crystal structure in the
space group of $\mathit{P}$4$\mathit{/m}$$\mathit{m}$$\mathit{m}$,
together with small amount impurities phases of RbFe$_{2}$As$_{2}$
(6.2\% wt) and EuFe$_{2}$As$_{2}$ (4.4\% wt), as shown in Fig. 3(a)
and 3(c).

In stark contrast to the magnetic satellite peaks displayed in RbEu(Fe$_{0.95}$Ni$_{0.05}$)$_{4}$As$_{4}$
arising from the helical magnetic structure of Eu, here at 2 K, well
below $\mathit{T_{m}}$, the magnetic scatterings due to the ordering
of Eu$^{2+}$ spins appear on top of the nuclear reflections for RbEu(Fe$_{0.91}$Ni$_{0.09}$)$_{4}$As$_{4}$,
which is shown in Fig. 3(e-g) for $\mathcal{\mathit{Q}}$
= (0 0 1) (e), (0 0 2) (f) and (0 0 3) (g) measured with a high resolution using
$\lambda$ = 2.41 \AA{}. This clearly indicates a magnetic propagation
vector of $\mathit{k}$ = 0.

Magnetic representation analysis for $\mathit{k}$ = 0 for the space
group of $\mathit{P}$4$\mathit{/m}$$\mathit{m}$$\mathit{m}$ yields
only two possible irreducible representations for the Eu(1$\mathit{a}$)
site (see the supplemental materials for details), labeled as $\Gamma_{8}$ and $\Gamma_{9}$, respectively. They
correspond to the collinear ferromagnetic structures in which all the
Eu$^{2+}$ moments are aligned along the $\mathit{c}$ axis and in
the $\mathit{ab}$ plane, respectively. Although no magnetization
data on single-crystal RbEu(Fe$_{0.91}$Ni$_{0.09}$)$_{4}$As$_{4}$
is available, the moment direction of Eu$^{2+}$ spins can still be
identified according to the nature of magnetic neutron diffraction.
As the magnetic scattering is only sensitive to the component of the
moment perpendicular to $\mathit{Q}$, dramatic enhancements of intensities
of (0 0 L) peaks and no visible changes of (H K 0) peak intensities
suggest that the Eu$^{2+}$ moments are mostly lying in the $\mathit{ab}$
plane so that the magnetic structure model described by $\Gamma_{8}$
can be excluded. Indeed the $\Gamma_{9}$ model with all spins aligned
along in-plane (1 1 0) direction fits the diffraction patterns at
2 K quite well, as shown by the solid curves in Fig. 3(b) and 3(d). As the fraction of the EuFe$_{2}$As$_{2}$ impurity phase is quite small (4.4\% wt), including the its magnetic phase in the refinement has no visible effect on the fitting of the 2 K data and the results about the 1144 main phase. Fixing the nuclear structure parameters and the scale factor derived
from the refinement of 20 K data, the refinement of 2 K data yields
the moment size of Eu to be 6.5(1) $\mathit{\mu_{B}}$ (see Table
1). Please note that a lower saturated moment of 6.0  $\mathit{\mu_{B}}$/Eu for $\mathit{x}$ = 0.09 in Ref. \onlinecite{Liu_17} is because of some nonmagnetic Eu$_{2}$O$_{3}$ impurities forming in older samples due to oxidation of metallic Eu. In fact, the saturated moment of Eu$^{2+}$ spins should be independent of the Ni doping level. The magnetic structure of RbEu(Fe$_{0.91}$Ni$_{0.09}$)$_{4}$As$_{4}$
is illustrated in Fig. 1(d). Compared with the undoped Eu1144 and
RbEu(Fe$_{0.95}$Ni$_{0.05}$)$_{4}$As$_{4}$ with $\mathit{x}$
= 0.05, the rotation angle between the moments in neighboring Eu layers
diminishes to zero for RbEu(Fe$_{0.91}$Ni$_{0.09}$)$_{4}$As$_{4}$
with $\mathit{x}$ = 0.09, forming a collinear in-plane ferromagnetic
structure. The magnetic space group of this helical structure is determined to be $\mathit{Cmm'm'}$ (No. 65.486).

\begin{figure}
\centering{}\includegraphics{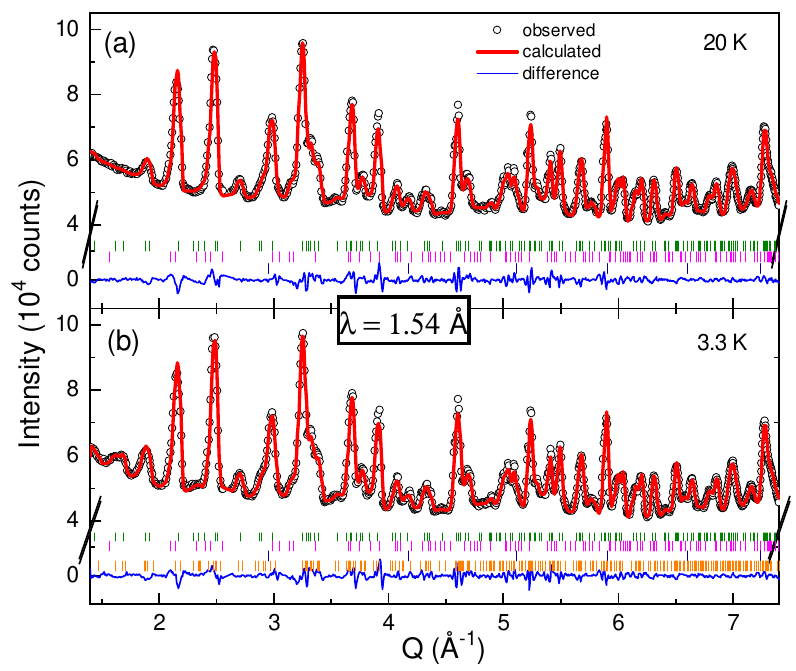}

\caption{NPD patterns of RbEu(Fe$_{0.93}$Ni$_{0.07}$)$_{4}$As$_{4}$ at
20 K (a) and 3.3 K (b) collected using the incident neutron wavelength
of 1.54 \AA{} and the Rietveld refinments. The pattern in (b) is the refinement
result obtained by adopting a magnetic structure model with the irreducible
representation $\Gamma_{5}$ as described in the text. The circles
represent the observed intensities, and the solid lines are the calculated
patterns. The differences between the observed and calculated intensities
are shown at the bottom. The vertical bars in olive, magenta, navy
and orange colors indicate the expected nuclear Bragg reflection from
RbEu(Fe$_{0.93}$Ni$_{0.07}$)$_{4}$As$_{4}$, FeAs impurity, vanadium
sample container and the magnetic Bragg reflection from RbEu(Fe$_{0.93}$Ni$_{0.07}$)$_{4}$As$_{4}$,
respectively. }
\end{figure}

After presenting the results of RbEu(Fe$_{1-x}$Ni$_{x}$)$_{4}$As$_{4}$ with $\mathit{x}$ = 0.05 and 0.09, we come to the magnetic structure determination of the SFM RbEu(Fe$_{0.93}$Ni$_{0.07}$)$_{4}$As$_{4}$ ($\mathit{T_{SC}}$ = 11.2 K) with $\mathit{T_{m}}$ and $\mathit{T_{SDW}}$ being 15.1 K and 35.0 K, respectively.\cite{Liu_17} As shown in Fig. 4(a), the diffraction pattern of RbEu(Fe$_{0.93}$Ni$_{0.07}$)$_{4}$As$_{4}$ at 20 K can be well fitted with the nuclear crystal structure in the space group of $\mathit{P}$4$\mathit{/m}$$\mathit{m}$$\mathit{m}$ together with a small amount impurities phase of FeAs (5.9\% wt). Again, no magnetic peaks at (0.5, 0.5, 3) arising from the Fe-AFM order can be identified. Upon cooling down to the base temperature of 3.3 K, the magnetic scattering due to magnetic ordering of Eu$^{2+}$ spins sets in. Unfortunately the high-resolution datasets with $\lambda$ = 2.41 \AA{} is lacking for this sample, due to the limited neutron beamtime. However, by setting the magnetic propagation vector $\mathit{k}$ itself as a variable parameter in the refinement of 3.3 K data, the diffraction pattern can be fitted pretty well with $\mathit{k}$ finally converged to (0, 0, 0.071(7)) and the moment size of Eu$^{2+}$ spins being 6.3(2) $\mathit{\mu_{B}}$, as shown in Table 1 and Fig. 4(b). This result corresponds to a helical magnetic structure similar to that of RbEu(Fe$_{0.95}$Ni$_{0.05}$)$_{4}$As$_{4}$, but with a smaller helix rotation angle of $\sim$ 26$^{\circ}$.

Using first-principles calculations, the energetic properties of different spin configurations of the Eu$^{2+}$ moments are computed for RbEuFe$_{4}$As$_{4}$ and RbEu(Fe$_{0.875}$Ni$_{0.125}$)$_{4}$As$_{4}$, respectively. As shown in Table 2 and 3, it is found that the noncollinear helical structure with $\mathit{k}$ = (0, 0, 0.25) possesses the lowest energy for the parent compound RbEuFe$_{4}$As$_{4}$, while the collinear ferromagnetic structure with the Eu$^{2+}$ moments lying in the $\mathit{ab}$ plane is energetically favorable for RbEu(Fe$_{0.875}$Ni$_{0.125}$)$_{4}$As$_{4}$ with $\mathit{x}$ = 0.125. These are well consistent with our experimental findings that the rotation angle between the moments in neighboring Eu layers diminishes with increasing Ni doping and the helical structure finally transforms into a purely collinear ferromagnetic structure.

\begin{table}
\caption{Energetic properties of the different spin configurations of the Eu$^{2+}$ moments for RbEuFe$_{4}$As$_{4}$. The results are the total energy difference per Eu atom. The helical, antiparallel and parallel configurations correspond to the magnetic structures in which the in-plane ferromagnetic Eu$^{2+}$ moments on adjacent layers are vertical, antiparallel, and parallel, respectively.}

\begin{ruledtabular} %
\begin{tabular}{cccccc}
configurations & $\Delta$$\mathit{E}$(meV) & $\mathit{M}_{Eu}$($\mu_{B}$)\tabularnewline
\hline 
helical ($\mathit{k}$ = (0, 0, 0.25)) & 0 & 6.986 \tabularnewline
antiparallel & 49.71 & 6.962 \tabularnewline
parallel & 49.21 & 6.962 \tabularnewline
\end{tabular}\end{ruledtabular}
\end{table}

\begin{table}
\caption{Energetic properties of the different spin configurations of the Eu$^{2+}$ moments for RbEu(Fe$_{0.875}$Ni$_{0.125}$)$_{4}$As$_{4}$. The results are the total energy difference per Eu atom. The helical, antiparallel and parallel configurations correspond to the magnetic structures in which the in-plane ferromagnetic Eu$^{2+}$ moments on adjacent layers are vertical, antiparallel, and parallel, respectively.}

\begin{ruledtabular} %
\begin{tabular}{cccccc}
configurations & $\Delta$$\mathit{E}$(meV) & $\mathit{M}_{Eu}$($\mu_{B}$)\tabularnewline
\hline 
helical ($\mathit{k}$ = (0, 0, 0.25)) & 0 & 6.971 \tabularnewline
antiparallel & 2.01 & 6.965 \tabularnewline
parallel & -2.04 & 6.970 \tabularnewline
\end{tabular}\end{ruledtabular}
\end{table}

\section{Discussion And Conclusion}

As shown in Fig. 1, the magnetic structure of the Eu$^{2+}$ moments in RbEu(Fe$_{1-x}$Ni$_{x}$)$_{4}$As$_{4}$ undergoes a smooth evolution from the helical structure, in which the in-plane ferromagnetically
aligned Eu$^{2+}$ spins on adjacent layers rotate by 90$^{\circ}$, gradually to a collinear ferromagnetic structure, in which all the Eu$^{2+}$ spins point along the tetragonal (1 1 0) direction. The $\mathit{c}$-component of the magnetic propagation vector, $\mathit{k_{z}}$, and the helix rotation angle ($\theta$) are plotted in Fig. 5(a) and 5(b) as a function of the Ni content $\mathit{x}$, respectively. Both of them diminish with increasing Ni content, in accordance with the gradual suppression of superconductivity as reported in Ref. \onlinecite{Liu_17}. 

\begin{figure}
\centering{}\includegraphics{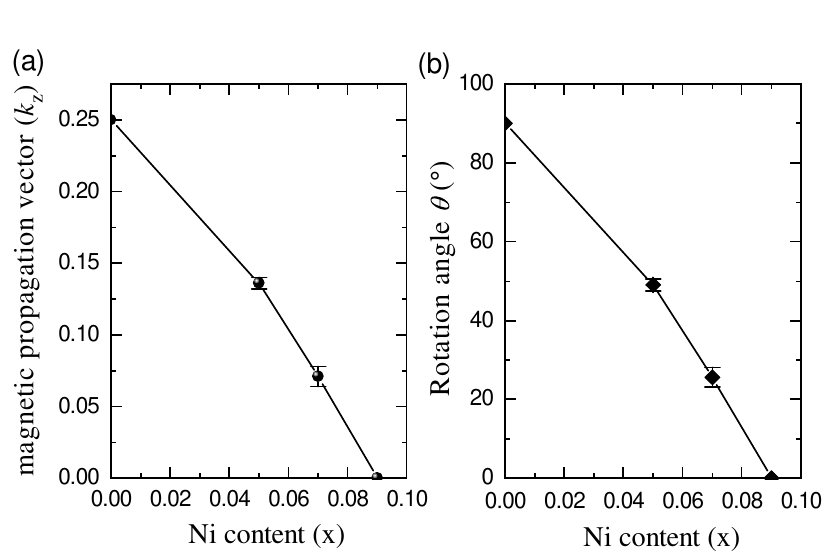}

\caption{The evolution of magnetic propagation vector $\mathit{k}$ = (0, 0,
$\mathit{k_{z}}$) (a) and the rotation angle ($\theta$) of the Eu$^{2+}$
spins between adjacent Eu layers (b) in RbEu(Fe$_{1-x}$Ni$_{x}$)$_{4}$As$_{4}$
as a function of the Ni content $\mathit{x}$.}
\end{figure}

\begin{figure*}
\centering{}\includegraphics[width=1\textwidth]{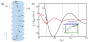}

\caption{An illustration of the spin directions in different layers of the helical magnetic structure (a) and a semi-quantitative description of the interlayer couplings as well as the rotation angle $\theta$ in the helix as a function of  $\mathit{k_{F}}r$ (b). The nearest ($\mathit{J_{c\mathrm{1}}}$, black solid line) and next-nearest ($\mathit{J_{c\mathrm{2}}}$, red solid line) interplayer couplings are assumed to be in the form of $\mathit{J_{c\mathrm{1}}}$ = $\mathit{C}$$\mathit{J_{\mathrm{RKKY}}}\mathrm{cos}(2k_{F}r)/r^{3}$ and $\mathit{J_{c\mathrm{2}}}$ = $\mathit{C}$$\mathit{J_{\mathrm{RKKY}}}\mathrm{cos}(4k_{F}r)/(2r)^{3}$, respectively, where $\mathit{C}$ is a scaling constant. The helix rotation angle $\theta$ (in the inset) is then calculated using $\mathit{\mathrm{cos}\theta}$${}=-\frac{\mathit{J_{c\mathrm{1}}}}{4\mathit{J_{c\mathrm{2}}}}$. The vertical dashed lines in (b) mark the possible $\mathit{k_{F}}r$ values of undoped Eu1144, where $\mathit{J_{c\mathrm{1}}}$ = 0 and $\mathit{J_{c\mathrm{2}}}$ \textless{} 0. Assuming that for undoped Eu1144 $\mathit{k_{F}}r$ = 2.5$\pi$, the black, red, and olive arrows next to the corresponding solid circles represent the shifts of $\mathit{J_{c\mathrm{1}}}$, $\mathit{J_{c\mathrm{2}}}$, and $\theta$ values with the decrease of $\mathit{k_{F}}r$ induced by Ni doping. The blue diamonds in the inset of (b) represent the $\theta$ values for different Ni content $\mathit{x}$ determined experimentally as shown in Fig. 5(b) for comparison.}
\end{figure*}

It was reported recently that in isostructural CaK(Fe$_{1-x}$Ni$_{x}$)$_{4}$As$_{4}$, the Ni doping may lead to the emergence of a hedgehog-type spin-vortex crystal (SVC) order of the Fe moments,\cite{Meier_18,Kreyssig_18} which is different from the stripe-type Fe-SDW order observed in \textquotedblleft 122\textquotedblright{} family iron pnictides.\cite{Dai_15,Su_09,Xiao_09} However, within our experimental resolution, the proposed recovery of Fe-AFM order with Ni doping can not be identified at $\mathit{Q}$ = (0.5, 0.5, $\mathit{L}$) ($\mathit{L}$ = integers), probably due to the weakness of related magnetic reflections from small Fe$^{2+}$ moments and high background in the NPD measurements. Future neutron diffraction experiments on large single-crystal samples of RbEu(Fe$_{1-x}$Ni$_{x}$)$_{4}$As$_{4}$, if available, will be crucial to confirm the possibly restored antiferromagnetism in the Fe sublattice.

The variation of the magnetic structure of Eu in RbEu(Fe$_{1-x}$Ni$_{x}$)$_{4}$As$_{4}$ can be understood semi-quantitatively in consideration of the exchange couplings. As the magnetism of Eu in Eu1144 is believed to be of a 2D character, the helix rotation angle $\theta$ between the ferromagnetic Eu$^{2+}$ layers predominantly depend on the competition between the nearest ($\mathit{J_{c\mathrm{1}}}$) and next-nearest ($\mathit{J_{c\mathrm{2}}}$) interplayer couplings (see Fig. 6(a)), with $\mathit{\mathrm{cos}\theta}=-\frac{\mathit{J_{c\mathrm{1}}}}{4\mathit{J_{c\mathrm{2}}}}$.\cite{Blundell} These exchange couplings between interlayer Eu$^{2+}$ moments is realized through the indirect Ruderman-Kittel-Kasuya-Yosida (RKKY) interaction $\mathit{J_{\mathrm{RKKY}}}$, mediated by the conduction $\mathit{d}$ electrons on the FeAs layers, in the form of $\mathit{J_{c}}\varpropto\mathit{J_{\mathrm{RKKY}}}\mathrm{cos}(2k_{F}r)/r^{3}$, where $\mathit{r}$ denotes the interlayer distance between the Eu$^{2+}$ moments and $\mathit{k_{F}}$ is the Fermi vector.\cite{Ruderman_54,Kasuya_56,Yosida_57,Akbari_13} Using first-principles calculations, it is figured out that the RKKY interaction strength $\mathit{J_{\mathrm{RKKY}}}$ is isotropic and barely changed upon Ni-doping ($\sim$ 0.12 meV).\cite{Xu_19} In the undoped Eu1144, $\mathit{J_{c\mathrm{1}}}$ is expected to be zero (for $\theta=90{^\circ}$ and $\mathit{\mathrm{cos}\theta}=0$), consistent with the 2D character of the Eu magnetism. This corresponds to $\mathit{\mathrm{2}k_{F\mathrm{0}}}r_{0}=(2n+1)\pi/2$, with $\mathit{k_{F\mathrm{0}}}$ and $\mathit{r_{\mathrm{0}}}$ being the Fermi vector and nearest interlayer distance between the Eu$^{2+}$ moments without Ni doping. $\mathit{J_{c\mathrm{2}}}\varpropto\mathit{J_{\mathrm{RKKY}}}\mathrm{cos}(4k_{F\mathrm{0}}r_{0})/(2r_{0})^{3}$ is therefore negative, responsible for the antiferromagnetic next-nearest interlayer coupling. As the hole carriers are compensated by the substitution of Ni$^{2+}$ (3$\mathit{d}$$^{8}$) for Fe$^{2+}$ (3$\mathit{d}$$^{6}$), the Fermi vector $\mathit{k_{F}}$ shrinks effectively, leading to the decrease of $\mathit{k_{F}}r$. Therefore, $\left|\mathit{J_{c\mathrm{1}}}\right|$ increases and $\left|\mathit{J_{c\mathrm{2}}}\right|$ decreases as the value of $\mathit{k_{F}}r$ is tuned away from $(\mathit{n}+1/2)\pi$, resulting in the increase of $\mathit{\mathrm{cos}\theta}$ ($=-\frac{\mathit{J_{c\mathrm{1}}}}{4\mathit{J_{c\mathrm{2}}}}$) and the decrease of $\theta$, as shown in Fig. 6(b) and its inset. This well explains the gradual disapperance of relative rotation between adjacent Eu$^{2+}$ layers with increasing Ni doping as determined experimentally.

It is argued that the emergence of helical magnetic structure with a period of four unit cells along the $\mathit{c}$ axis in undoped Eu1144 ($\mathit{k}$ = (0, 0, 0.25)) is favored by the exchange interaction
between superconductivity and ferromagnetism,\cite{Devizorova_19} as predicted by Anderson and Suhl long time ago to be one solution for the compromise between these two antagonistic phenomena.\cite{Anderson_59} As an alternate scenario, it is proposed theoretically that the ferromagnetic contribution to the interlayer RKKY interaction from the non-superconducting normal parts and the antiferromagnetic contribution from the superconducting layers compete with each other, giving rise to the helical ground-state magnetic configuration as a result of frustration.\cite{Koshelev_19} It is worth pointing out that our experimental results are also qualitatively consistent with these arguments. On one hand, the helix rotation angle $\theta$ diminishes with Ni doping, thus releasing the frustration in favor of a collinear ferromagnetic structure. On the other hand, according to the prediction by Anderson and Suhl, the periodicity of the spin helix $\mathit{d}$ is correlated with the supercondcuting coherence length ${\xi_\mathrm{0}}$ in the form of $\mathit{d}\varpropto{(\xi_\mathrm{0})^{1/3}}$.\cite{Anderson_59} As the superconducting transition temperature $\mathit{T_{SC}}$ and the upper critical field $\mathit{H_{c\mathrm{2}}}$ decrease with increasing Ni doping,\cite{Willa_20} ${\xi_\mathrm{0}}$ increases according to the Ginzburg-Landau formalism $\mathit{H_{c\mathrm{2}}=\mathrm{\Phi_{0}}}/2\pi{\xi_\mathrm{0}}^2$, which is consistent with the diminishing $\theta$ and increasing helix periodicity $\mathit{d}$. Although some recent spectroscopic measurements seem to suggest the decoupling of magnetism from Eu from superconducting FeAs layers,\cite{Hemmida_21,Kim_21} we note that a recent scanning Hall microscopy experiment has revealed a pronounced suppression of the superfluid density near the Eu magnetic ordering temperature in Eu1144, indicating a pronounced exchange interaction between the superconducting and magnetic subsystems.\cite{Collomb_21}

In conclusion, the magnetic structures of RbEu(Fe$_{1-x}$Ni$_{x}$)$_{4}$As$_{4}$ superconductors are systematically investigated by neutron powder diffraction. It is found that as the superconductivity gets suppressed gradually with the increase of Ni doping, the magnetic propagation vector of the Eu sublattice diminishes, corresponding to the decrease of the rotation angle between the moments in neighboring Eu layers with a helical structure. For non-superconducting RbEu(Fe$_{0.91}$Ni$_{0.09}$)$_{4}$As$_{4}$, all the Eu$^{2+}$ spins point along the tetragonal (1 1 0) direction, forming a purely collinear ferromagnetic structure. Such an evolution from helical to collinear ferromagnetic order of the Eu$^{2+}$ spins with increasing Ni doping is well supported by first-principles calculations. The variation of the rotation angle between adjacent Eu$^{2+}$ layers can be well explained by considering the change of magnetic exchange couplings mediated by the indirect RKKY interaction. 

\bibliographystyle{apsrev} \bibliographystyle{apsrev}
\begin{acknowledgments}
This work is partly based on experiments performed at the Australian
Nuclear Science and Technology Organisation (ANSTO), Sydney, Australia
and the Institut Laue-Langevin (ILL), Grenoble, France. W.T.J. would like to acknowledge
Shang Gao, Hao Deng and Karen Friese for helpful discussions. The authors
acknowledge the supports by the National Natural Science Foundation of China (Grant No. 12074023 and 11927807), the National Key Research and Development
Program of China (2016YFA0300202), the Fundamental Research Funds
for the Central Universities in China (YWF-20-BJ-J-1043 and YWF-21-BJ-J-1044), and Shanghai Science and Technology Committee (Grant Nos. 19ZR1402600 and 20DZ1100604). 

\bibliographystyle{apsrev}
\bibliography{1144}
\end{acknowledgments}

\end{document}